\documentclass[3p,times,procedia]{elsarticle}

\usepackage{nuphb_ecrc}
\usepackage{amsmath}
\usepackage{caption}

\volume{00}
\firstpage{1}
\journalname{Nuclear Physics B}
\runauth{F.\ Capozzi et al.}
\jid{nuphb}
\jnltitlelogo{Nuclear Physics B}
\biboptions{sort&compress}

\begin{document}
\begin{frontmatter}

\title{Neutrino masses and mixings: \\ Status of known and unknown $3\nu$ parameters}

\author[s1,s2]{F.~Capozzi}
\address[s1]{Dipartimento di Fisica e Astronomia ``Galileo Galilei'', Universit\`a di Padova, Via F.\ Marzolo 8, I-35131 Padova, Italy}
\address[s2]{Istituto Nazionale di Fisica Nucleare (INFN), Sezione di Padova, Via F.\ Marzolo 8, I-35131 Padova, Italy}
\ead{francesco.capozzi@pd.infn.it}

\author[s3]{E.~Lisi}
\address[s3]{Istituto Nazionale di Fisica Nucleare (INFN), Sezione di Bari, Via E.\ Orabona 4, I-70126 Bari, Italy}
\ead{eligio.lisi@ba.infn.it}

\author[s4,s3]{A.~Marrone}
\address[s4]{Dipartimento Interateneo di Fisica ``Michelangelo Merlin'', Universit\`a di Bari, Via G.\ Amendola 173, I-70126 Bari, Italy}
\ead{antonio.marrone@ba.infn.it}

\author[s5,s6]{D.~Montanino}
\address[s5]{Dipartimento di Matematica e Fisica ``Ennio de Giorgi'', Universit\`a del Salento, Via Arnesano, I-73100 Lecce, Italy}
\address[s6]{Istituto Nazionale di Fisica Nucleare (INFN), Sezione di Lecce,  Via Arnesano, I-73100 Lecce, Italy}

\ead{daniele.montanino@le.infn.it}

\author[s4,s3]{A.~Palazzo}
\ead{antonio.palazzo@ba.infn.it}

\begin{abstract}
Within the standard $3\nu$ mass-mixing framework, 
we present an up-to-date global analysis of neutrino oscillation data (as of January 2016), 
including the latest available results from experiments with atmospheric neutrinos (Super-Kamiokande and IceCube DeepCore),
at accelerators (first T2K $\overline\nu$  and NO$\nu$A $\nu$ runs in both
appearance and disappearance mode), and at short-baseline reactors (Daya Bay and RENO far/near spectral ratios), 
as well as a reanalysis
of older KamLAND data in the light of the ``bump'' feature recently observed in reactor spectra.   
We discuss improved constraints on the five known oscillation parameters ($\delta m^2,\,|\Delta m^2|,\,\sin^2\theta_{12},\,\sin^2\theta_{13},\sin^2\theta_{23}$), and the status of the three remaining unknown parameters:  
the mass hierarchy [sign$(\pm\Delta m^2)$], the $\theta_{23}$ octant [sign$(\sin^2\theta_{23}-1/2)$], and the
possible CP-violating phase $\delta$. With respect to previous global fits,
we find that the reanalysis of KamLAND data induces a slight decrease of
both $\delta m^2$ and $\sin^2\theta_{12}$, while the latest accelerator and atmospheric data induce a slight increase
of $|\Delta m^2|$. Concerning the unknown parameters, we confirm 
the previous intriguing preference for negative values of $\sin\delta$ 
(with best-fit values around $\sin\delta \simeq -0.9$), but we find no statistically significant indication 
about the $\theta_{23}$ octant or the mass hierarchy (normal or inverted). 
Assuming an alternative (so-called LEM) analysis
of NO$\nu$A data, some $\delta$ ranges can be excluded at $>3\sigma$, and the
normal mass hierarchy appears to be slightly favored at $\sim\! 90\%$ C.L.  We also describe in detail
the covariances of selected pairs 
of oscillation parameters. Finally,
we briefly discuss the implications of the above results
on the three non-oscillation observables sensitive to the (unknown) absolute $\nu$ mass scale: the sum of $\nu$
masses $\Sigma$ (in cosmology), the effective $\nu_e$ mass $m_\beta$ (in beta decay),
and the effective Majorana mass $m_{\beta\beta}$ (in neutrinoless double beta decay).
\end{abstract}

\end{frontmatter}

\section{Introduction}

Yellow, blue and dark blue: this is the simple color palette used for painting and penning each of the two-sided Nobel Diplomas awarded to Takaaki Kajita \cite{Dip1} and Arthur B.\ McDonald \cite{Dip2}. On the left side, one can gaze at an artist's view---sketched with a few broad strokes---of the neutrino transformative trip from the bright yellow Sun, through the Earth's blue darkness, into a blue pool of water  \cite{Ulla}. On the right side, one can read the---beautifully and precisely penned---Nobel laureate names and prize motivations, in ink colors that continuously change
from deep blue to blue with yellow shades \cite{Anni}.
 In a sense, the two sides of the Diplomas evoke the interplay between a broad-brush picture of $\nu$ masses and mixings (the pioneering era) and carefully designed measurements and theoretical descriptions (the precision era), 
in a continuous feedback between breakthrough and control, that may open the field to further fundamental discoveries 
\cite{Lect}.

In this paper, we aim at presenting both the broad-brush features and the fine structure
of the current picture of neutrino oscillation phenomena, involving 
the mixing of the three neutrino states having definite flavor $\nu_{e,\mu,\tau}$ with three states  $\nu_{1,2,3}$ 
having definite masses $m_i$ \cite{PDGR}. Information on known and unknown neutrino mass-mixing parameters is derived by 
a global analysis of neutrino oscillation data, that extends and
updates our previous work \cite{Ours} with recent experimental inputs, as discussed in Sec.~2  
(see also \cite{Gonz,Vall} for previous global analyses by other groups). In Sec.~3,
precise constraints (at few \% level) are obtained on four well-known oscillation parameters, namely, 
the squared-mass differences  $\delta m^2=m^2_2-m^2_1$ and $\Delta m^2=m^2_3-(m^2_1+m^2_2)/2$, and the mixing angles $\theta_{12}$ and $\theta_{13}$. Less precise constraints, including an octant ambiguity, are reported for the angle $\theta_{23}$. In this picture, we also
discuss the current unknowns related to the neutrino mass hierarchy [sign$(\Delta m^2)$]  and to the possible leptonic CP-violating phase $\delta$. The trend favoring negative values of $\sin\delta$ appears to be confirmed, with best-fit values around $\delta \simeq 1.3$--$1.4\,\pi$ (i.e., $\sin\delta\simeq -0.9$). More fragile indications, which depend on alternative analyses of specific data sets, concern the exclusion of some $\delta$ ranges at $>3\sigma$, and a slight preference for normal hierarchy at 90\% C.L. The covariances of selected parameter pairs, and the implications for non-oscillation searches, are presented in Sec.~4 and 5, respectively.  Our conclusions are summarized in Sec.~6.

\section{Global analysis: methodology and updates}

In this section we discuss methodological issues and input updates for the global analysis. Readers interested  
only in the fit results may jump to Sec.~3. 
 
In general, no single oscillation experiment can currently probe, with high sensitivity, the full parameter space spanned by 
the mass-mixing variables $(\delta m^2,\,\pm\Delta m^2,\,\theta_{12},\,\theta_{13},\,\theta_{23},\,\delta)$. One can then
group different data sets, according to their specific sensitivities or complementarities with respect
to some oscillation parameters. We follow the methodology of Refs.~\cite{Ours,Prev} as summarized below.

We first combine the data coming from solar and KamLAND reactor experiments (``Solar+KL'') with those coming from
long-baseline accelerator searches in both appearance and disappearance modes (``LBL Acc''). The former data set constrains the $(\delta m^2,\,\theta_{12})$ parameters (and, to some extent, also $\theta_{13}$ \cite{Ours,Prev}), which are a crucial 
input for the $3\nu$ probabilities relevant to the latter data set. The combination ``LBL Acc+Solar+KL data'' provides 
both upper and lower bounds on the $(\delta m^2,\,|\Delta m^2|,\,\theta_{12},\,\theta_{13},\,\theta_{23})$ parameters but,
by itself, is not particularly sensitive to $\delta$ or to sign($\pm\Delta m^2$) ($+$ for normal hierarchy, NH, and $-$ for 
inverted hierarchy, IH).  

The LBL Acc+Solar+KL data are then combined with short-baseline reactor data (``SBL Reactors''), that provide strong constraints on the $\theta_{13}$ mixing angle via disappearance event rates, as well as on useful bounds on
$\Delta m^2$ via spectral data (when available). The synergy between LBL Acc+Solar+KL data and SBL Reactor data increases significantly the sensitivity to $\delta$ \cite{Ours}.      

Finally, we add atmospheric neutrino data (``Atmos''), which probe both flavor appearance and disappearance channels for $\nu$ and $\overline\nu$,  both in vacuum and in matter,  with a  very rich phenomenology  
spanning several decades in energy and path lengths. This data set is dominantly sensitive to the mass-mixing pair ($\Delta m^2,\,\theta_{23}$) and, subdominantly, to all the
other oscillation parameters. Despite their complexity, atmospheric  data may thus add useful pieces of information 
on subleading effects (and especially on the three unknown parameters), 
which may either support or dilute the indications coming 
from the previous data sets. 

In all cases, the fit results are obtained by minimizing a $\chi^2$ function, that depends on the arguments
$(\delta m^2,\,\pm\Delta m^2,\,\theta_{12},\,\theta_{13},\,\theta_{23},\,\delta)$ and on a number of systematic nuisance 
parameters via the pull method \cite{Ours,Pull}. Allowed parameter ranges at $N_\sigma$ standard deviations are defined via $N^2_\sigma=
\chi^2-\chi^2_\mathrm{min}$ \cite{PDGR}. The same definition is maintained in covariance plots involving parameter pairs, so that 
the previous $N_\sigma$ ranges 
are recovered by projecting the allowed regions onto each axis. Undisplayed parameters are marginalized away.

A final remark is in order. The definition $N^2_\sigma = \chi^2 -\chi^2_{\mathrm{min}}$ is based on Wilks' theorem \cite{PDGR}, that is not strictly applicable to discrete choices (such as NH vs IH, see \cite{Qian,Patt,Stan} and references 
therein) or to cyclic variables (such as $\delta$, see \cite{Schw,Baye}). Concerning hierarchy tests, it has been argued that the 
above $N_\sigma$ prescription can still be used to 
assess the statistical difference between NH and IH with good approximation \cite{Colo}. 
Concerning CP violation tests, the prescription appears to lead (in general) to more conservative bounds on $\delta$, as compared with the results  obtained from numerical experiments \cite{Blenn,Elev,Gonz}. 
In principle, one can construct the correct $\chi^2$ distribution by generating extensive replicas of all the relevant data sets via Monte Carlo simulations, randomly spanning
the space of the neutrino oscillation and systematic nuisance parameters. However, such a construction 
would be extremely time-consuming and  is
beyond the scope of this paper. 
For the sake of simplicity, we shall  
adopt the conventional $N_\sigma$ definition, supplemented by cautionary comments when needed.  

\subsection{Solar + KL data analysis and the reactor spectrum bump}

With respect to \cite{Ours}, the solar neutrino analysis is unchanged. Concerning KamLAND (KL) reactor neutrinos, 
we continue to use the 2011 data release \cite{KL11} as in \cite{Ours}. We remark that the latest published KL data \cite{KL13} are divided into three subsets, with correlated systematics that are difficult to implement outside the Collaboration.%
\footnote{It would useful to release future KL data in a format allowing more reproducible analyses.} 
In this work, we reanalyze the 2011 KL data for the following reason. 

The KL analysis requires the (unoscillated) absolute reactor $\overline\nu_e$ spectra as input. In this context,
a new twist has been recently provided by the observation of a $\sim\!10\%$ event excess in the range
 $E_\nu\sim 5$--7~MeV (the so-called ``bump'' or ``shoulder'') \cite{Bump}, with respect to the expectations from reference 
Huber-M\"uller (HM) spectra \cite{Hube,Mull},
in each of the current high-statistics SBL reactor experiments RENO \cite{RENO}, 
Double Chooz \cite{DCHO} and Daya Bay \cite{DBay}. 

This new spectral feature is presumably due to nuclear physics effects 
(see the recent review in \cite{Nove}), whose origin is still subject to investigations and debate 
\cite{Dwye,Zaka,Haye,Jaff,Sonz}. In principle, one would like to know in detail the separate spectral modifications for each reactor fuel component \cite{Hase,Voge}. However,
the only information available at present is the overall energy-dependent ratio $f(E)$ between data and HM predictions, which
we extract (and smooth out) from the latest Daya Bay results (see Fig.~3 in \cite{DBay}). 

We use the $f(E)$ ratio as an effective fudge factor multiplying 
the unoscillated HM spectra for KamLAND, which are thus anchored to the absolute  Daya Bay spectrum  \cite{DBay}. 
In our opinion, this overall correction can capture the main bump effects in the KL spectral analysis. 
More refined KL data fits will be possible when the bump feature(s) will be better 
understood and broken down into separate spectral components.
Concerning the KL dominant oscillation parameters $(\delta m^2,\,\theta_{12})$, we find that the inclusion of the bump fudge factor induces a slight negative shift of their best-fit values, which persists in combination
with solar data (see Sec.~3).

Finally, we recall that the $3\nu$ analysis of solar+KL data is performed in terms of three free parameters $(\delta m^2,\,\theta_{12},\,\theta_{13})$, providing a weak but interesting indication for nonzero $\theta_{13}$ \cite{Ours,Prev,Hint}. Tiny differences between  transition probabilities in NH and IH  \cite{Quas,Glob} are negligible within the present accuracy. The  hierarchy-independent function $\chi^2(\delta m^2,\,\theta_{12},\,\theta_{13})$, derived from the solar+KL data fit, is then used in combination with the following LBL accelerator data. 

\subsection{LBL Accelerator data analysis}

With respect to \cite{Ours}, we include the most recent results from the  Tokai-to-Kamioka (T2K) experiment in Japan and 
from the NO$\nu$A experiment at Fermilab, in both appearance and disappearance modes. In particular, we include 
the latest T2K neutrino data \cite{T2KN} and the first T2K antineutrino data \cite{T2KA}, as well as the first NO$\nu$A 
neutrino data as of January 2016 \cite{NOV1,NOV2}. The statistical analysis of LBL experiments has been performed using a modified version of the
software GLoBES \cite{GLO1,GLO2} for the calculation of the expected number of events.%
\footnote{
In disappearance mode we have fitted energy spectra, which constrain $\Delta m^2$ and $\theta_{23}$ via the
position and amplitude of the oscillation dip, respectively. In appearance mode, characterized by much lower statistics, 
we have fitted the total number of events.  We have checked that, even for T2K $\nu$ appearance data, total-rate or spectral analyses of events produce very similar results in the global fit. 
}
 For each LBL data set, the $\chi^2$ function takes into account Poisson statistics \cite{PDGR} and the main systematic 
 error sources, typically related to energy-scale errors and to normalization uncertainties of signals  and  
 backgrounds, as taken
from \cite{T2KN,T2KA,NOV1,NOV2}. Concerning NO$\nu$A $\nu_e$ appearance data, the collaboration used two different event selection methods for increasing the purity of the event sample: A primary method based on a likelihood identification (LID) selector, and a secondary one based on a library event matching (LEM) selector, leading to somewhat different results for the $\nu_e$ signal and background \cite{NOV1}. We shall consider the LID data as a default choice for NO$\nu$A, but we shall also comments on the 
impact of the alternative LEM data.  
 
 We have reproduced with good approximation the allowed parameter regions shown by 
 T2K \cite{T2KN,T2KA} and by NO$\nu$A \cite{NOV2} (in both LID and LEM cases \cite{NOV1}), under the same hypotheses or restrictions 
 adopted therein for the
undisplayed parameters. We remark that, in our global analysis (see Sec.~4), all the oscillation parameters are left unconstrained. Note that we define the parameter $\Delta m^2$, driving the dominant 
LBL oscillations, as 
\begin{equation}
\label{DM2}
\Delta m^2=m^2_3-(m^2_1+m^2_2)/2\ ,
\end{equation}
in both NH ($\Delta m^2>0$) and IH ($\Delta m^2<0$) \cite{Quas,Glob}.
A comparative discussion of this and alternative conventions in terms of $\Delta m^2_{13}$, $\Delta m^2_{23}$, $\Delta m^2_A$, $\Delta m^2_{\mu\mu}$ and $\Delta m^2_{ee}$ is reported in \cite{Bil1,Bil2} and references therein. Although any such convention is immaterial (as far as the full $3\nu$ oscillation probabilities are used), the adopted one must be explicitly declared, since 
the various definitions differ by terms of $O(\delta m^2)$, comparable to
the current $\pm1\sigma$ uncertainty of $\Delta m^2$.

\subsection{SBL Reactor data analysis}

With respect to \cite{Ours}, we include herein the spectral data on the far-to-near detector ratio as a function of energy, as recently reported by the experiments  Daya Bay  (Fig.~3 of \cite{DB15}) and RENO (Fig.~3 of \cite{RENO}). Besides the statistical errors, we include a simplified set of pulls for energy-scale and flux-shape systematics, since the bin-to-bin correlations are not publicly reported in \cite{DB15,RENO}. We neglect systematics related to the spectral bump feature, which affect absolute spectra (see Sec.~2.1) but largely cancel in the analysis of far/near ratios (see \cite{DB15,RENO}).

We reproduce with good accuracy the joint allowed regions reported in \cite{DB15} and \cite{RENO} for the mixing amplitude $\sin^22\theta_{13}$ and their effective squared mass parameters $\Delta m^2_{ee}$, for both NH and IH.%
\footnote{For a recent discussion and comparison of different $\Delta m^2_{ee}$ definitions
and conventions, see \protect\cite{Par1}. In any case, in our global fits 
we always use the $\Delta m^2$ parameter defined  in Eq.~(\protect\ref{DM2}).
}
	We then combine the Daya Bay and RENO analyses, in terms of our default parameters $\sin^2\theta_{13}$ and $\Delta m^2$. The combined fit 
results are dominated, for both mass-mixing parameters, by the high-statistics Daya Bay data. While the reactor bounds on
$\theta_{13}$ are extremely strong,  the current bounds on $\Delta m^2$ are not yet competitive with those coming from LBL accelerator data in disappearance mode, although they help in reducing slightly its uncertainty (see Sec.~4).

\subsection{Atmospheric $\nu$ data analysis}

With respect to \cite{Ours}, we update our analysis of Super-Kamiokande (SK) atmospheric neutrino data by including the latest (phase I-IV) data as taken from \cite{Wend,Wen1}. We also include for the first time the recent atmospheric data released by
the IceCube DeepCore (DC) collaboration \cite{Deep,DCwb,Yane}. We reproduce with good accuracy the joint bounds on the
$\sin^2\theta_{23}$ and $\Delta m^2_{32}$ parameters shown by DC in \cite{Deep}, under the same assumptions used therein. 
In this work, the $\chi^2$ functions for SK and DC have been simply added. In the future, it may be useful to isolate and
properly combine possible systematics which may be common to SK and DC 
(related, e.g., to flux and cross section normalizations). 

\vspace*{1mm}
\section{Global $3\nu$ analysis: constraints on single oscillation parameters}

In this Section we discuss the constraints on known and unknown  oscillation parameters, 
coming from the global $3\nu$ analysis of all the data discussed above. The impact of different
data sets will be discussed in the next Section.

\begin{figure}[t]\centering
  \includegraphics[width=0.575\textwidth]{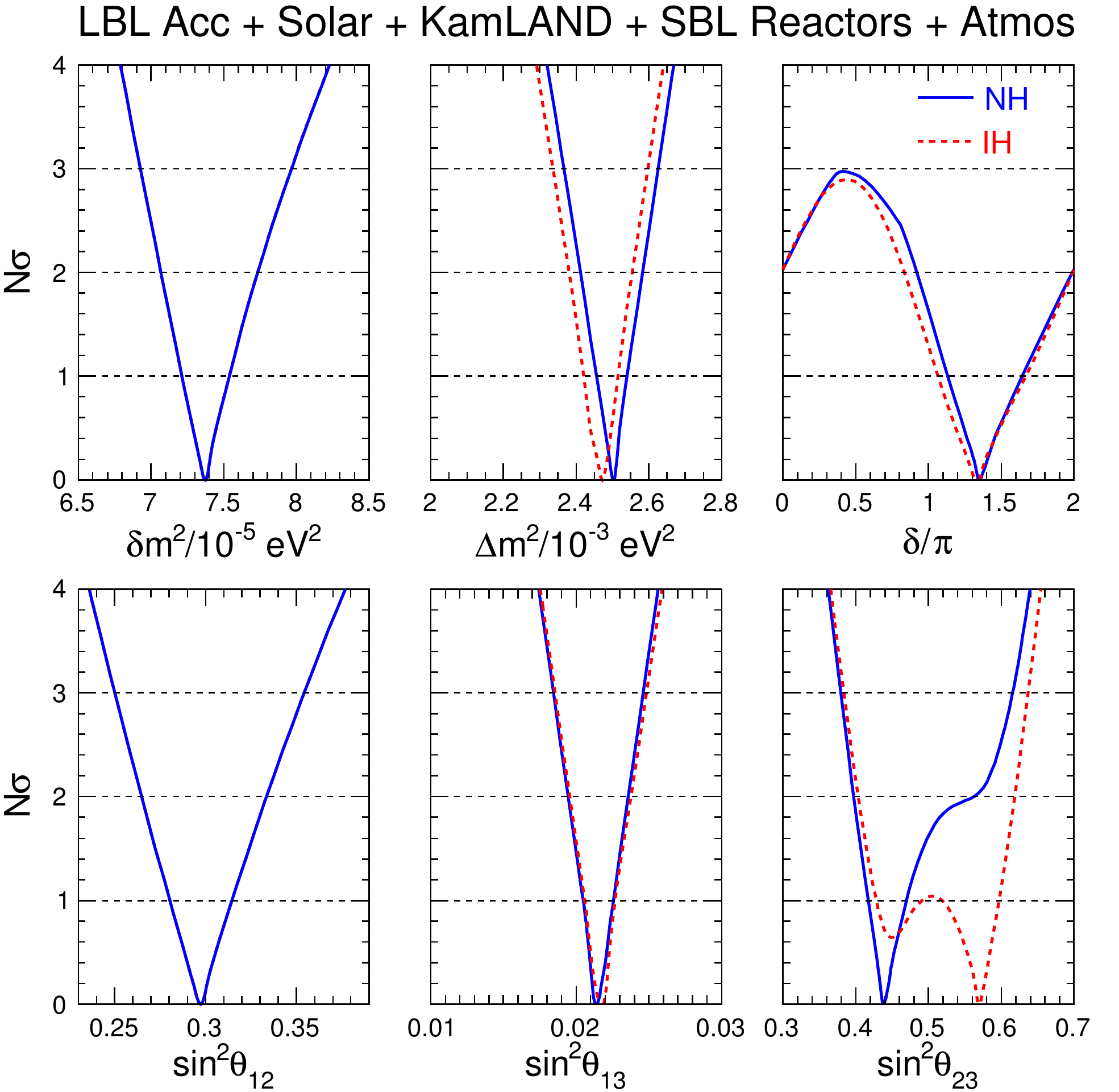}
  \caption{\small Global analysis of neutrino oscillation data. Bounds on the mass-mixing parameters are given in 
  terms of standard deviations $N_\sigma$ from the best fit, for either NH (solid lines) or IH (dashed lines).
  Bounds on $(\delta m^2,\,\sin^2\theta_{12})$ are hierarchy-independent. Horizontal
  dotted lines mark the 1, 2, and $3\sigma$ levels for each parameter.
   \label{fig01}}
\end{figure}

Figure~1 shows the bounds on single oscillation parameters, in terms of standard deviations $N_\sigma$ from the best fit.
Linear and symmetric curves would correspond to gaussian uncertainties---a situation realized with excellent approximation for
the $(\Delta m^2,\,\sin^2\theta_{13})$ mass-mixing pair and, to a lesser extent, for the $(\delta m^2,\,\sin^2\theta_{12})$
pair. The best fit of the $\sin^2\theta_{23}$ parameter flips from the first to the second octant by changing the hierarchy 
from normal to inverted, but this indication is not statistically significant, since maximal mixing $(\sin^2\theta_{23}=1/2)$ is 
allowed at $\sim\!1.6\sigma$ ($\sim\!90\%$~C.L.) for NH and at $\sim 1\sigma$ for IH. In any case, all these parameters 
 have both upper and lower bounds well above the $3\sigma$ level. If we define the average $1\sigma$ error 
as $1/6$ of the $\pm3\sigma$ range, our global fit implies the following fractional
uncertainties: $\delta m^2$ (2.4\%), $\sin^2\theta_{12}$ (5.8\%), $\Delta m^2$ (1.8\%), $\sin^2\theta_{13}$ (4.7\%), and
$\sin^2\theta_{23}$ (9\%).  

The parameter $\delta$ is associated to a Dirac phase in the neutrino mixing matrix, which
might induce leptonic CP violation effects for $\sin \delta\neq 0$ \cite{PDGR}. Recent fits to global $\nu$ data 
\cite{Ours,Gonz,Vall} and partial (LBL accelerator) data \cite{Elev,Pala}
have consistently shown a preference for negative values of $\sin\delta$, as a 
result of the combination of LBL accelerator $\nu$ and $\overline\nu$ data and
 of SBL reactor data. The reason is that the LBL
appearance probability  contains a CP-violating part proportional to $-\sin\delta$ ($+\sin\delta$)
for neutrinos (antineutrinos) \cite{PDGR}. With respect to the CP-conserving case $\sin\delta=0$, 
values of $\sin\delta <0$  are then expected to produce a slight increase (decrease) of events in $\nu_\mu\to\nu_e$ 
($\overline\nu_\mu\to\overline\nu_e$) oscillations for $\theta_{13}$ fixed (by reactors), consistently with   
the appearance results of T2K (using both $\nu$ \cite{T2KN} and $\overline\nu$ \cite{T2KN}) and  in NO$\nu$A (using $\nu$ \cite{NOV1}), although within  large statistical uncertainties.%
\footnote{As already noted in \protect\cite{Ours}, the older MINOS LBL accelerator data (included in our global analysis) 
favor instead $\sin\delta >0$ \protect\cite{MINO} 
but with relatively low statistical significance, so that the overall preference for $\sin\delta <0$ from T2K
and NO$\nu$A is not spoiled.} 
 This trend for $\delta$ is clearly confirmed by the results in Fig.~1, which show a best fit for
 $\sin\delta \simeq -0.9$ ($\delta\simeq 1.3$--$1.4\pi$)
 in both NH and IH, while opposite values around $\sin\delta \simeq+0.9$ are disfavored at almost $3\sigma$ level. 
Although all values of $\delta$ are still allowed at $3\sigma$, the emerging indications in favor of 
$\sin\delta<0$ are intriguing and deserve further studies in T2K and NO$\nu$A, as well as in
future LBL accelerator facilities. We remark that our bounds on $\delta$ are conservative, and
that dedicated constructions of the $\chi^2$ distributions via extensive numerical simulations
might lead to stronger indications on $\delta$, as discussed  in Sec.~2.

\begin{table}[h]\centering
\captionsetup{width=.78\textwidth}
\caption{\label{Synopsis} \small Results of the global $3\nu$ oscillation analysis, in terms of best-fit values and
allowed 1, 2 and $3\sigma$ ranges  for the $3\nu$ mass-mixing parameters. See also Fig.~1 for a graphical representation 
of the results. We recall that $\Delta m^2$ is defined
as $m^2_3-{(m^2_1+m^2_2})/2$, with $+\Delta m^2$ for NH and $-\Delta m^2$ for IH. The CP violating phase is 
taken in the (cyclic) interval $\delta/\pi\in [0,\,2]$. The last row reports the (statistically insignificant) overall $\chi^2$ difference between IH and NH.}\vspace*{-1.5mm}
\resizebox{.8\textwidth}{!}{
\begin{tabular}{lccccc}
\hline\hline
Parameter & Hierarchy & Best fit & $1\sigma$ range & $2\sigma$ range & $3\sigma$ range \\
\hline
$\delta m^2/10^{-5}~\mathrm{eV}^2 $ & NH or IH & 7.37 & 7.21 -- 7.54 & 7.07 -- 7.73 & 6.93 -- 7.97 \\
\hline
$\sin^2 \theta_{12}/10^{-1}$ & NH or IH & 2.97 & 2.81 -- 3.14 & 2.65 -- 3.34 & 2.50 -- 3.54 \\
\hline
$\Delta m^2/10^{-3}~\mathrm{eV}^2 $ & NH & 2.50 & 2.46 -- 2.54 & 2.41 -- 2.58 & 2.37 -- 2.63 \\
$\Delta m^2/10^{-3}~\mathrm{eV}^2 $ & IH & 2.46 & 2.42 -- 2.51 & 2.38 -- 2.55 & 2.33 -- 2.60 \\
\hline
$\sin^2 \theta_{13}/10^{-2}$ & NH & 2.14 & 2.05 -- 2.25 & 1.95 -- 2.36 & 1.85 -- 2.46 \\
$\sin^2 \theta_{13}/10^{-2}$ & IH & 2.18 & 2.06 -- 2.27 & 1.96 -- 2.38 & 1.86 -- 2.48 \\
\hline
$\sin^2 \theta_{23}/10^{-1}$ & NH & 4.37 & 4.17 -- 4.70 & 3.97 -- 5.63 & 3.79 -- 6.16 \\
$\sin^2 \theta_{23}/10^{-1}$ & IH & 5.69 & 4.28 -- 4.91 $\oplus$ 5.18 -- 5.97 & 4.04 -- 6.18 & 3.83 -- 6.37 \\
\hline
$\delta/\pi$ & NH & 1.35 & 1.13 -- 1.64 & 0.92 -- 1.99  &  0 -- 2   \\
$\delta/\pi$ & IH & 1.32 & 1.07 -- 1.67 & 0.83 -- 1.99  &  0 -- 2  \\
\hline
$\Delta \chi^2_{\mathrm{{I}-{N}}}$ & IH$-$NH & +0.98 \\[1pt]
\hline\hline
\end{tabular}
}
\end{table}

Table~1 shows the same results of Fig.~1 in numerical form, with three significant digits for each parameter. 
In the last row of the table we add
a piece of information not contained in Fig.~1, namely, the $\Delta\chi^2$ difference between normal and inverted 
hierarchy. The NH is slightly favored over the IH at the (statistically insignificant) level
of $\sim 1\sigma$ in the global fit. We remark that both Fig.~1 and Table~1 use the NO$\nu$A LID data set in appearance mode
(see Sec.~2.2).

By adopting the alternative NO$\nu$A LEM data set, we find no variation for the 
$\delta m^2$ and $\sin^2\theta_{12}$ parameters (dominated  by Solar+KL data) and for the $\Delta m^2$ parameter
(dominated by LBL data in disappearance mode, in combination with atmospheric and reactor spectral data). 
We find slight variations for $\sin^2\theta_{13}$ and $\sin^2\theta_{23}$, and a small but interesting increase of 
the bounds on $\delta$ above the $3\sigma$ level. Figure~2 shows the corresponding results for the $\sin^2\theta_{23}$
and $\delta$ parameters, to be compared with the rightmost panels of Fig.~1.
   
\begin{figure}[hb]\centering
  \includegraphics[width=0.22\textwidth]{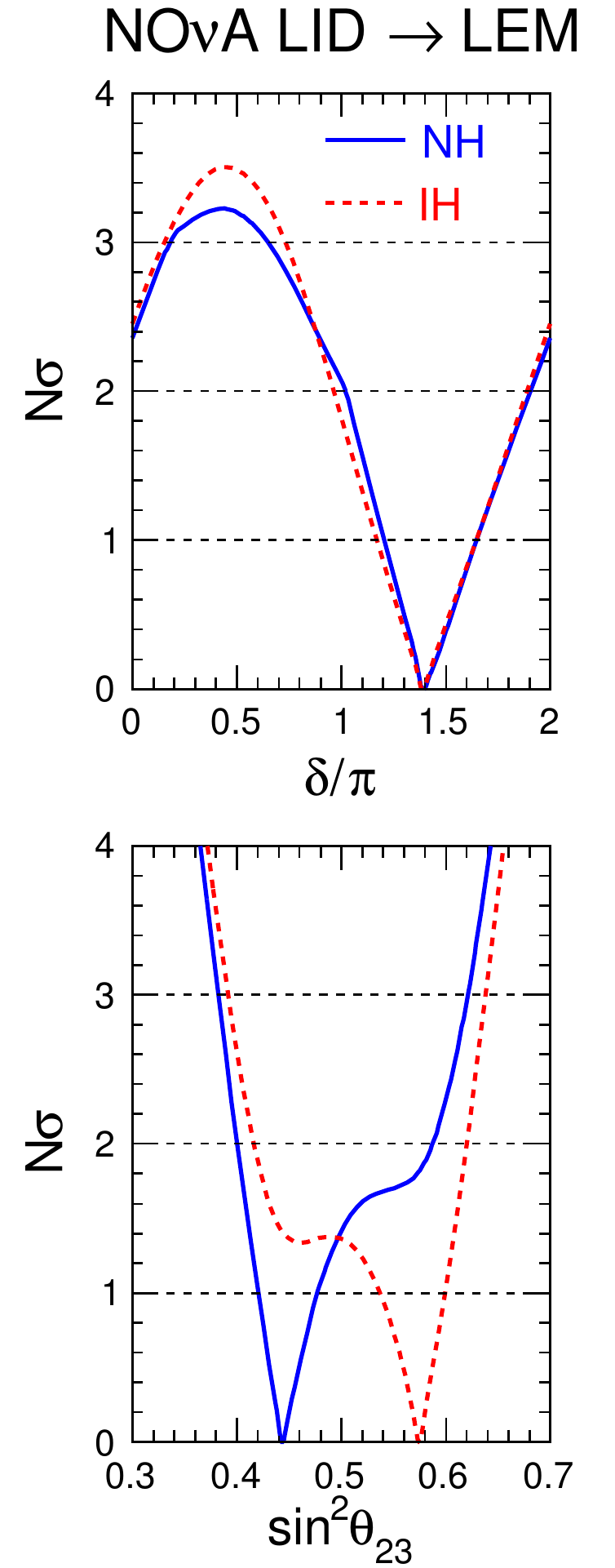}
  \caption{\small As in Fig.~1 (rightmost panels) but with NO$\nu$A LEM data replacing LID data. See the text for details.
  \label{fig02}}
\end{figure}

In Table~2 we report the results of the global fit using NO$\nu$A LEM data, but only for those parameters bounds which
differ from Table~1. Some intervals surrounding $\delta\simeq \pi/2$ can be excluded at $>3\sigma$. We also
find an increased sensitivity to the hierarchy, with the NH slightly favored (at $90\%$ C.L.) over the IH. 
These indications, although
still statistically limited, deserve some attention, for reasons that will be discussed in more detail at the end of the next
Section.

\begin{table}[t]\centering
\captionsetup{width=.78\textwidth}
\caption{\label{Synopsis2} \small As in Table~1, but using the
NO$\nu$A LEM (rather than LID) data set. Variations of the $\delta m^2$, $\theta_{12}$ and $\Delta m^2$ parameters 
(not shown) are numerically insignificant. The ranges $\delta/\pi\in [0.18,\,0.65]$ (NH) 
and $\delta/\pi\in [0.15,\,0.73]$ (IH) are disfavored at $>3\sigma$. See also Fig.~2 for a graphical representation
of the results for $\theta_{23}$ and $\delta$.
The NH is slightly preferred over the IH at $\sim\!90\%$~C.L.\ 
($\Delta \chi^2_{\mathrm{{I}-{N}}}=+2.8$).}\vspace*{-1.5mm}
\resizebox{.8\textwidth}{!}{
\begin{tabular}{lccccc}
\hline\hline
Parameter & Hierarchy & Best fit & $1\sigma$ range & $2\sigma$ range & $3\sigma$ range \\
\hline
$\sin^2 \theta_{13}/10^{-2}$ & NH & 2.17 & 2.06 -- 2.27 & 1.96 -- 2.37 & 1.86 -- 2.47 \\
$\sin^2 \theta_{13}/10^{-2}$ & IH & 2.19 & 2.08 -- 2.28 & 1.98 -- 2.38 & 1.88 -- 2.49 \\
\hline
$\sin^2 \theta_{23}/10^{-1}$ & NH & 4.43 & 4.21 -- 4.77 & 4.00 -- 5.87 & 3.82 -- 6.21 \\
$\sin^2 \theta_{23}/10^{-1}$ & IH & 5.75 & 5.37 -- 5.99 & 4.16 -- 6.20 & 3.92 -- 6.38 \\
\hline
$\delta/\pi$ & NH & 1.39 & 1.21 -- 1.65 & 1.02 -- 1.91  &  0 -- 0.18 $\oplus$ 0.65 -- 2  \\
$\delta/\pi$ & IH & 1.39 & 1.17 -- 1.64 & 0.96 -- 1.89  &  0 -- 0.15 $\oplus$ 0.73 -- 2 \\
\hline
$\Delta \chi^2_{\mathrm{{I}-{N}}}$ & IH$-$NH & +2.8 \\[1pt]
\hline\hline
\end{tabular}
}
\end{table}

\section{Global $3\nu$ analysis: Parameter covariances and impact of different data sets}

In this section we show and interpret the joint $N_\sigma$ contours (covariances) for selected pairs of oscillation 
parameters. We also discuss the impact of different data sets on such bounds.

We start with the analysis of the ($\delta m^2,\,\sin^2\theta_{12},\,\sin^2\theta_{13})$ parameters, which
govern the oscillations phenomenology of solar and KamLAND neutrinos.  Figure~3 shows the corresponding
bounds derived by a fit to Solar+KL data only (solid lines). By themselves, these data provide a $\sim\!1\sigma$
hint of $\theta_{13}>0$ \cite{Hint}, with a best fit ($\sin^2\theta_{13}\simeq 0.16$) 
close to current SBL reactor values.

\begin{figure}[bh]\centering
  \includegraphics[width=0.53\textwidth]{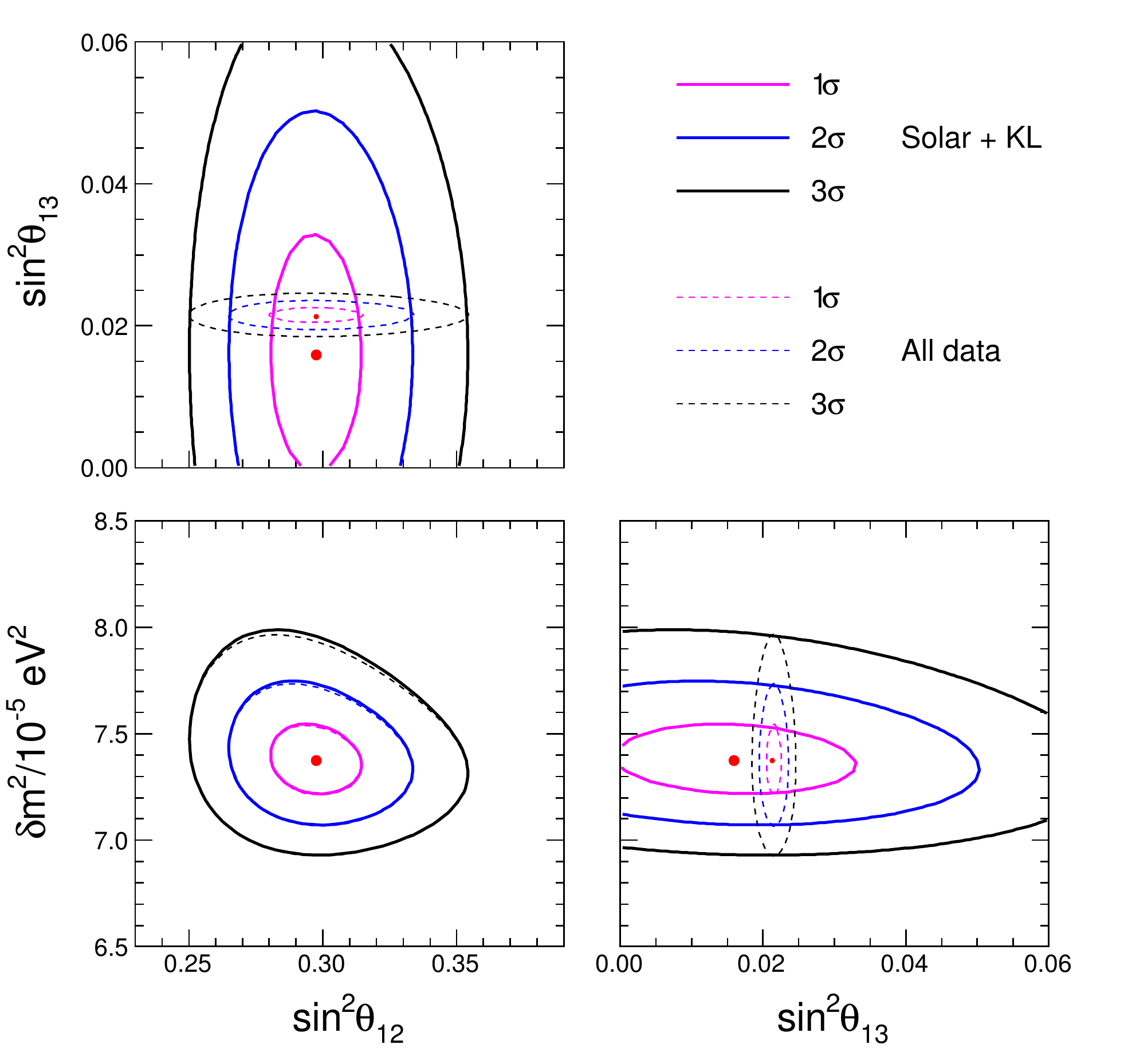}
  \caption{\small Bounds at $N_\sigma=1$, 2 and 3 on each pair of parameters chosen among ($\delta m^2,\,\sin^2\theta_{12},\,\sin^2\theta_{13})$, as derived by our analysis of Solar+KL data (solid lines) and of all data (dashed lines). 
 The dots mark the best-fit points. The bounds refer to NH case, and are very similar for IH case (not shown).     
  \label{fig03}}
\end{figure}

The $(\delta m^2,\,\sin^2\theta_{12})$ parameters in Fig.~3 appear to be slightly
anticorrelated, with a best-fit point at ($7.37\times 10^{-5}$~eV$^2$,$\,0.297$). These values are slightly
lower than those reported in our previous work ($7.54\times 10^{-5}$~eV$^2$,$\,0.308$) \cite{Ours},
as a result of altering the absolute KL spectra to account for the bump feature (see Sec.~2.1).
Statistically, these deviations amount to about $-1\sigma$ for $\delta m^2$ and $-0.6\sigma$ for $\sin^2\theta_{12}$,
and thus are not entirely negligible. A better understanding of the absolute reactor spectra (in both normalization and shape) 
is thus instrumental 
to analyze the KamLAND data with adequate precision.
 
Finally, Fig.~3 shows the joint bounds on the ($\delta m^2,\,\sin^2\theta_{12},\,\sin^2\theta_{13})$ parameters
from the global fit including all data (dashed lines). 
The bounds on the pair $(\delta m^2,\,\sin^2\theta_{12})$ are basically
unaltered, while those on $\sin^2\theta_{13}$ are shrunk by more than an order of magnitude, mainly as a result
of SBL reactor data.

Let us consider now the interplay between $\sin^2\theta_{13}$ and the mass-mixing parameters $\Delta m^2$ and $\sin^2\theta_{23}$, which dominate
the oscillations of LBL accelerator neutrinos.  Figure~4 shows the covariance plot for the $(\sin^2\theta_{13},\,\Delta m^2)$ 
parameters. Starting from the leftmost panels, one can see that the LBL Acc.+Solar+KL data, 
by themselves, provide both upper and lower bounds on $\sin^2\theta_{13}$ at $3\sigma$
level. The  
best-fit values of $\sin^2\theta_{13}$ lie around $\sim\!0.02$ 
in either NH or IH, independently of SBL reactor data. The best-fit 
values of $\Delta m^2$ are slightly higher than in our previous work \cite{Ours}, mainly as a result of the recent
NO$\nu$A data. The joint $(\sin^2\theta_{13},\,\Delta m^2)$ contours appear to be somewhat bumpy, as a result of the
octant ambiguity discussed below.%
\footnote{
In a sense, the allowed regions in Fig.~4 (and in the following covariance plots) can be considered as the union of two overlapping subregions, associated to  the quasi-degenerate octants of $\theta_{23}$.
}
 In the middle panels, the inclusion of SBL reactor data improves dramatically 
the bounds on $\sin^2\theta_{13}$ and, to a small but nonnegligible extent, also those on $\Delta m^2$. Finally, in the
rightmost panels, atmospheric data induce a small increase of its
central value (mainly as a result of DeepCore data), and a further 
reduction of the $\Delta m^2$ uncertainty. In comparison with \cite{Ours}, the $\Delta m^2$
value is shifted by $\sim\! 1\sigma$ upwards in the global fit.

Figure~5 shows the covariance plot for the $(\sin^2\theta_{23},\,\sin^2\theta_{13})$ 
parameters. The leftmost panels show a slight negative correlation and degeneracy between these two variables, 
which is induced by the dominant dependence of the LBL appearance channel on the product
$\sin^2\theta_{13}\sin^2\theta_{23}$, as 
also discussed in \cite{Ours,Prev}.  The overall LBL acc+Solar+KL preference for relatively low values of 
$\sin^2\theta_{13}$ ($\sim \!0.02$) breaks such a degeneracy and leads to a weak preference for the second octant.

\begin{figure}[bh]\centering
  \includegraphics[width=0.77\textwidth]{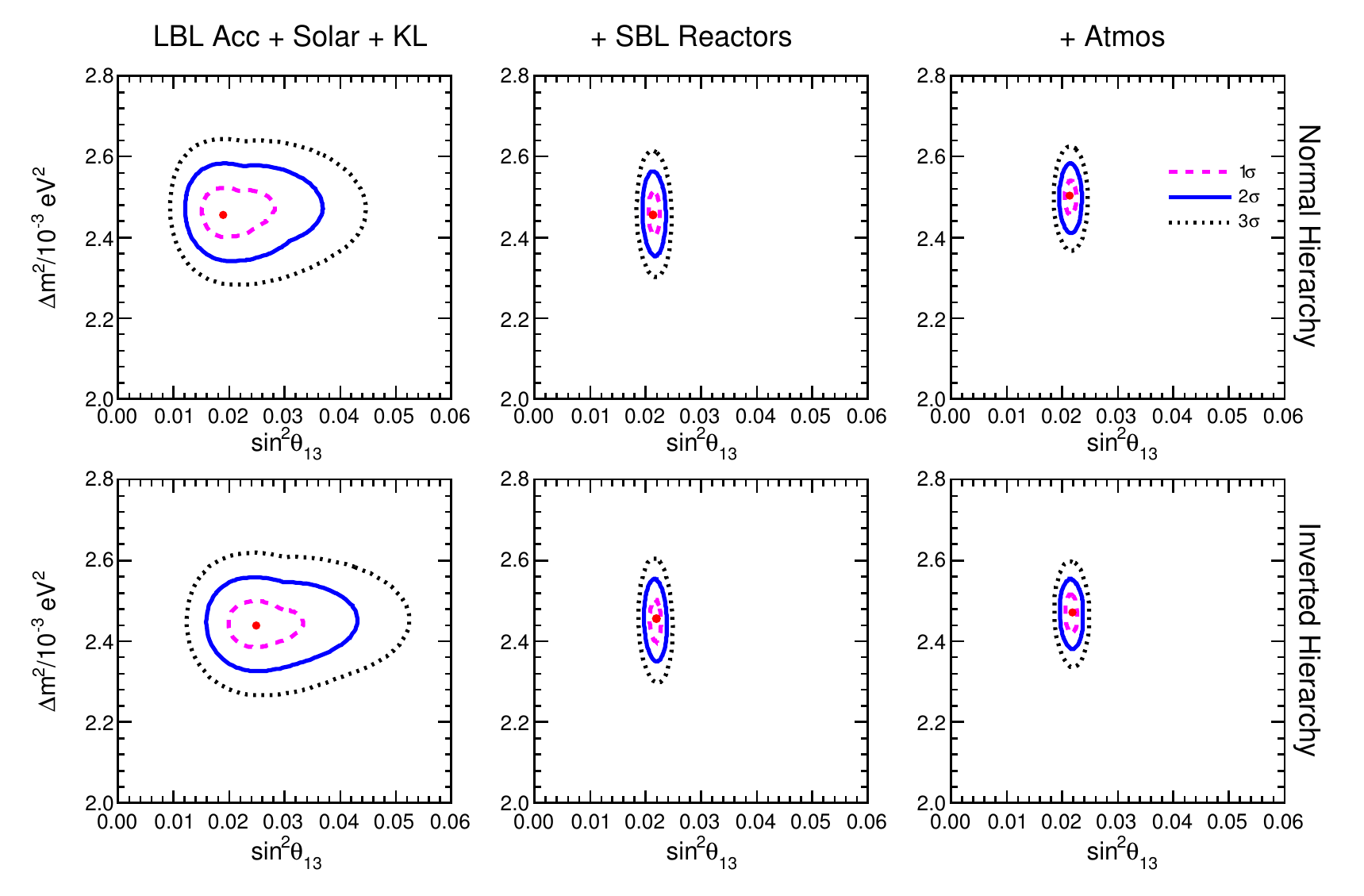}
  \caption{\small Covariance plot for the $(\Delta m^2,\,\sin^2\theta_{13})$ parameters. From left to right, the regions
  allowed at $N_\sigma=1$, 2 and 3 refer to the analysis of LBL Acc+Solar+KL data (left panels), plus SBL reactor data
  (middle panels), plus Atmospheric data (right panels), with best fits marked by dots. The three upper (lower) panels
  refer to NH (IH).   
  \label{fig04}}
\end{figure}

\begin{figure}[t]\centering
  \includegraphics[width=0.77\textwidth]{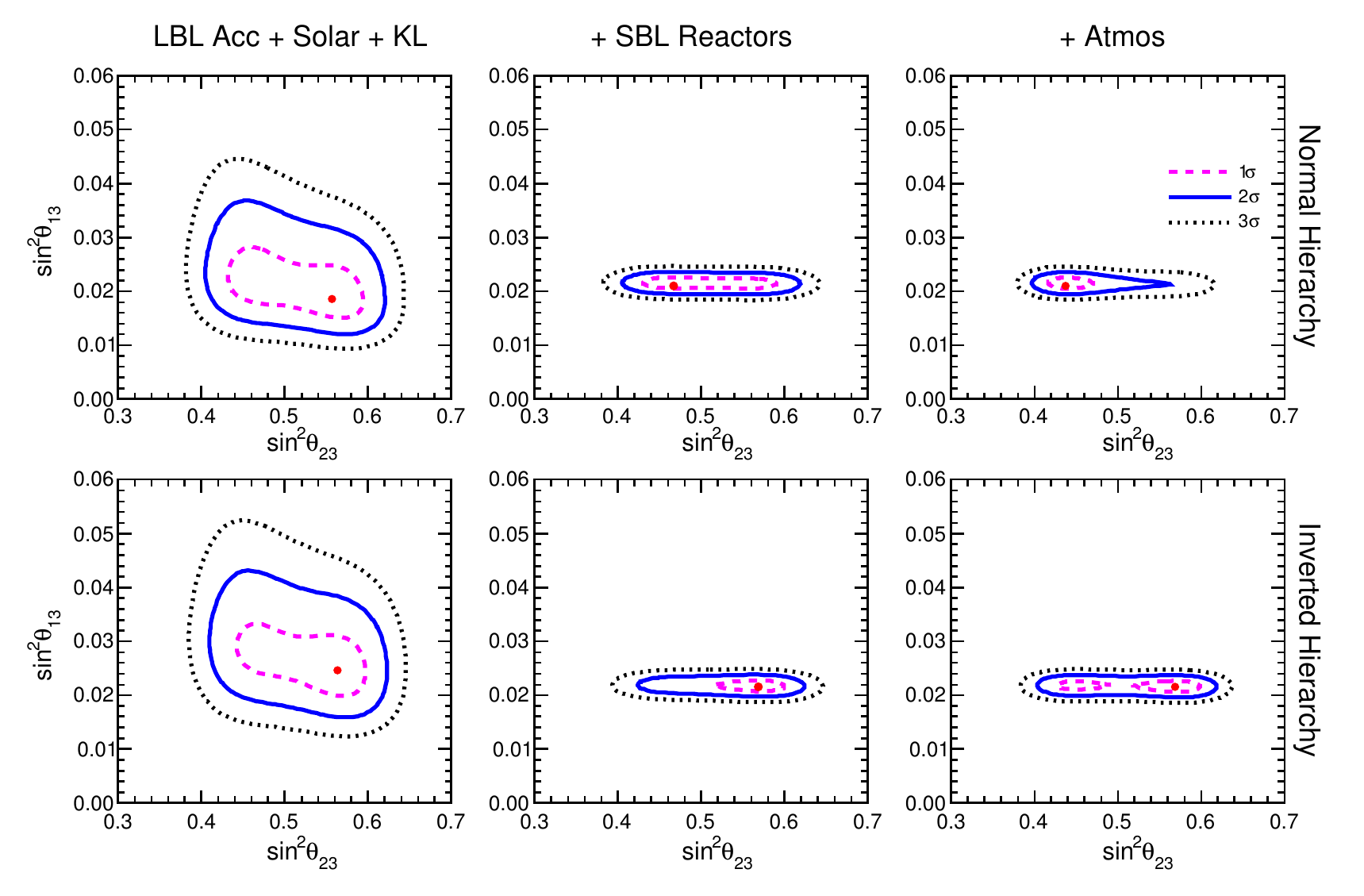}\vspace*{-3mm}
  \caption{\small As in Fig.~4, but for the $(\sin^2\theta_{23},\,\sin^2\theta_{13})$ parameters.  
  \label{fig05}}
\end{figure}
\vspace*{-5mm}

SBL reactor data (middle panels of Fig.~5) shrink the $\sin^2\theta_{13}$ range for both NH and IH. For IH, however, they
do not significantly change the central value of $\sin^2\theta_{13}$, nor the correlated  best-fit value of
$\sin^2\theta_{23}$, which stays in
the second octant. Conversely, for NH, the SBL reactor data do shift the central value of $\sin^2\theta_{13}$ upwards 
(with respect to the left panel), and best-fit value of
$\sin^2\theta_{23}$ is correspondingly shifted into 
first octant. Finally, the inclusion of atmospheric data (rightmost panels) alters the $N_\sigma$ contours, but
does not change the qualitative preference for the first (second) octant of $\theta_{23}$ in NH (IH).

Figure~6 shows the octant ambiguity in terms of bounds on the
mass-mixing parameters $(\Delta m^2,\,\sin^2\theta_{23})$. The fragility of current 
octant indications stems from the data themselves rather than on analysis details: nearly maximal mixing is
preferred by T2K (accelerator) and DeepCore (atmospheric) data, while nonmaximal mixing 
is preferred by MINOS and NO$\nu$A (accelerator) and by SK (atmospheric) data.
The combined results on $\theta_{23}$ appear thus still fragile, as far as the long-standing
octant degeneracy \cite{Octa} is concerned.

\begin{figure}[hb]\centering
  \includegraphics[width=0.77\textwidth]{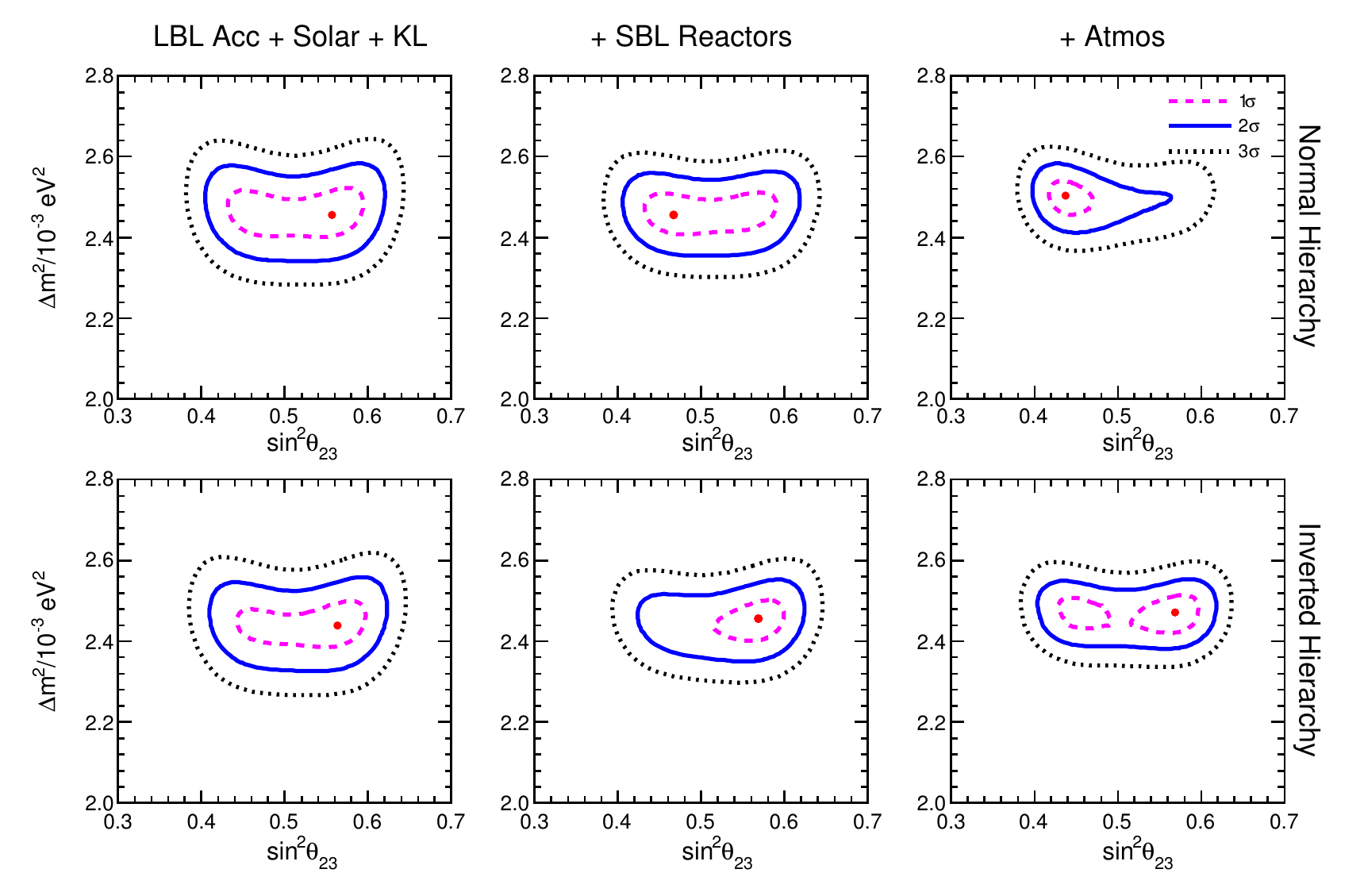}\vspace*{-3mm}
  \caption{\small As in Fig.~4, but for the $(\Delta m^2,\,\sin^2\theta_{23})$ parameters.  
  \label{fig06}}
\end{figure}

\begin{figure}[t]\centering
  \includegraphics[width=0.77\textwidth]{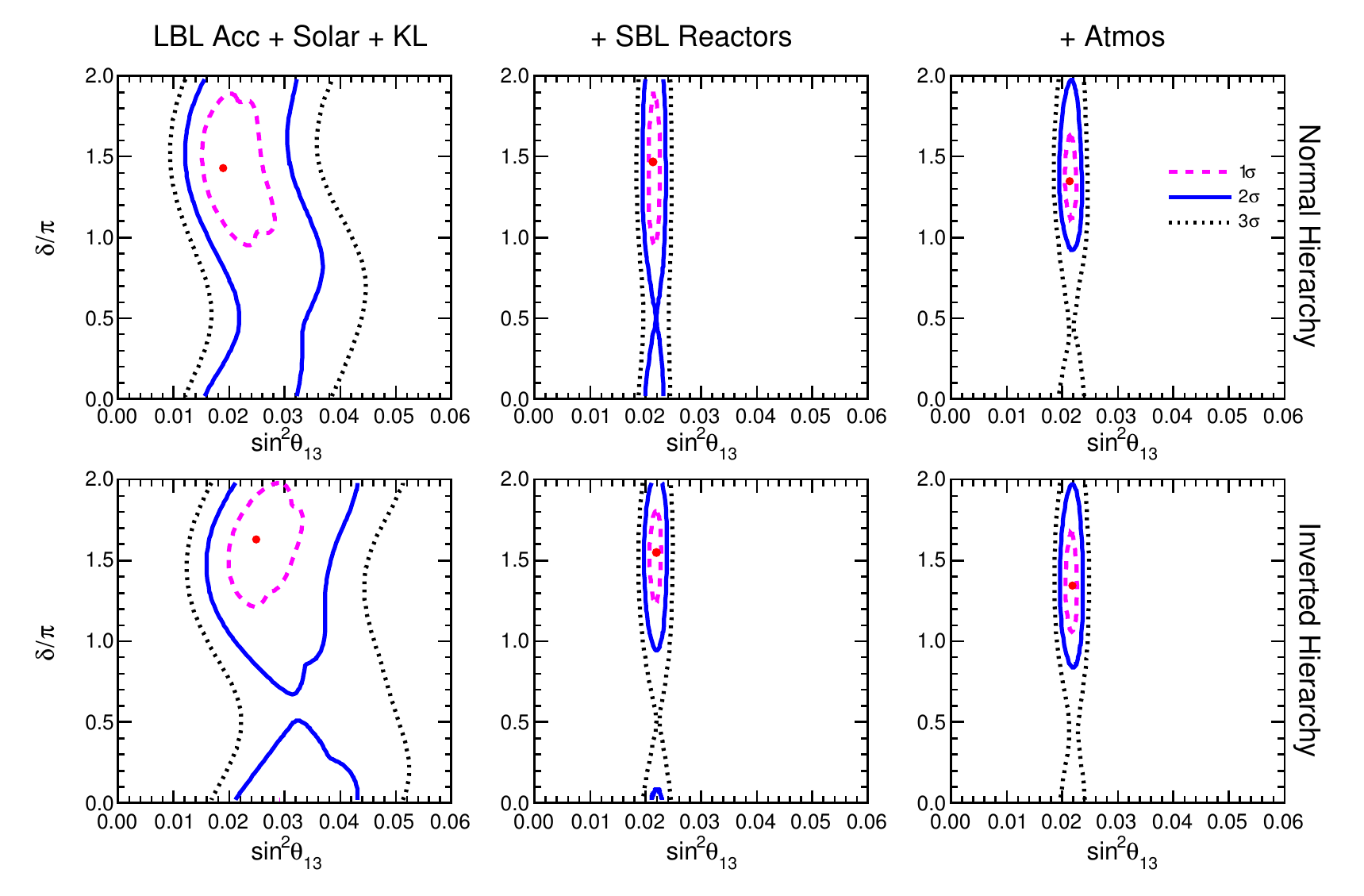}\vspace*{-3mm}
  \caption{\small As in Fig.~4, but for the $(\sin^2\theta_{13},\,\delta)$ parameters.  
  \label{fig07}}
\end{figure}

A very recent example of the (non)maximal $\theta_{23}$ issue 
is provided by the NO$\nu$A data in disappearance mode, which entailed
a preference for maximal mixing with preliminary data 
\cite{NOV0} and for nonmaximal mixing with definitive data \cite{NOV2}. We trace 
this change to the migration of a few events among reconstructed energy bins in the final 
NO$\nu$A data (not shown).

Let us complete the covariance analysis by discussing the interplay of the CP-violating phase 
$\delta$ with the mixing parameters $\sin^2\theta_{13}$ and $\sin^2\theta_{23}$. Figure~7 shows
the $N_\sigma$ bounds in the $(\sin^2\theta_{13},\,\delta)$ plane, which is at the focus
of LBL accelerator searches in appearance mode \cite{T2KA,NOV1,MINO}. The leftmost panels
show the wavy bands allowed by LBL Acc.+Solar+KL data, with a bumpy structure due to
the octant ambiguity (which was even more evident in older data fits \cite{Prev}). 
In the middle panels, SBL Reactor data select a narrow vertical strip, which does not alter 
significantly the preference  for $\delta\sim 3\pi/2$ stemming from
LBL Acc.+Solar+KL data alone. 

\begin{figure}[bh]\centering
  \includegraphics[width=0.77\textwidth]{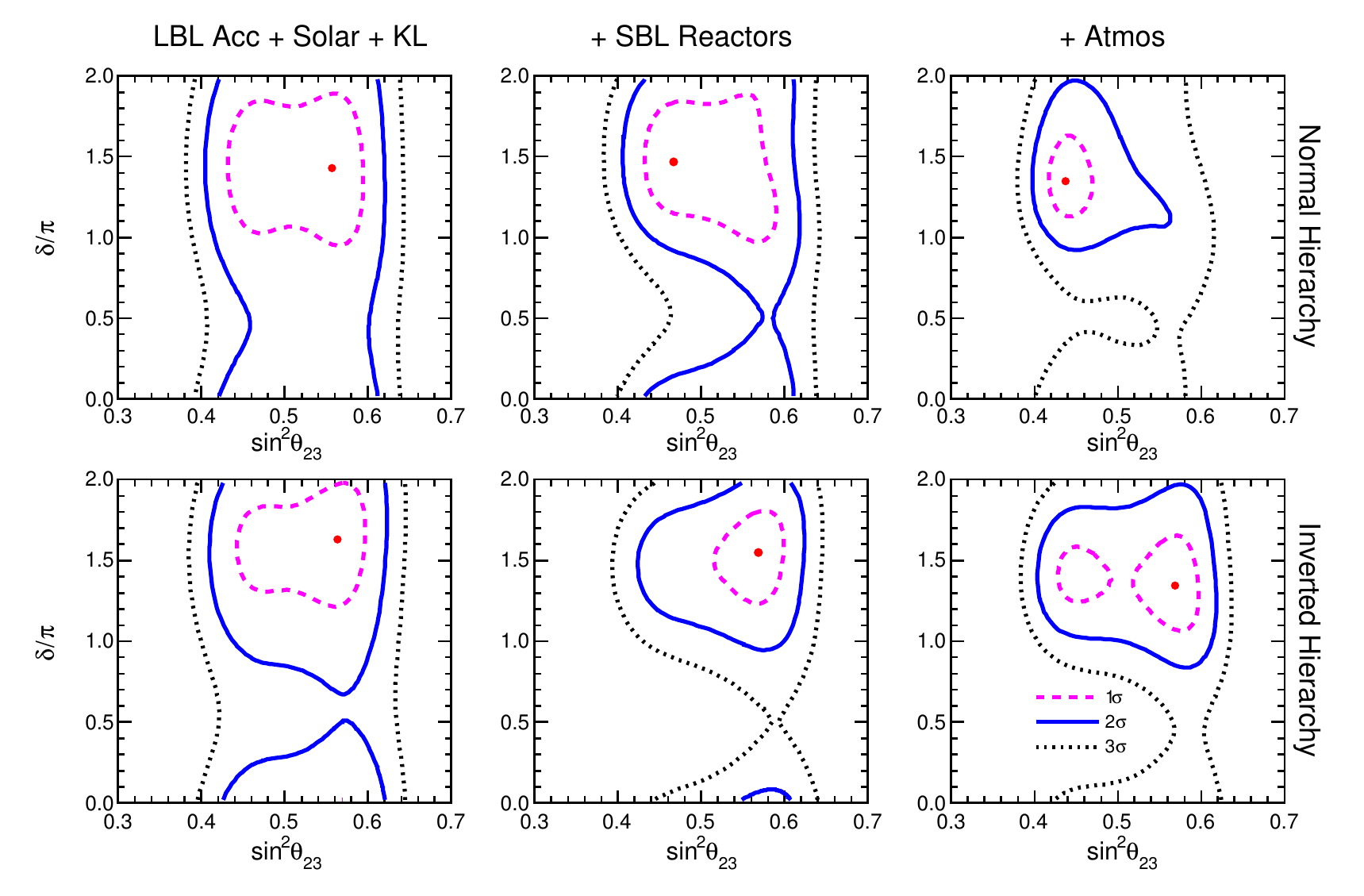}\vspace*{-3mm}
  \caption{\small As in Fig.~4, but for the $(\sin^2\theta_{23},\,\delta)$ parameters.  
  \label{fig08}}
\end{figure}

In this context, it is sometimes asserted that
the current preference for $\delta\sim 3\pi/2$ emerges from a ``tension'' between LBL accelerator and SBL reactor
data on $\theta_{13}$; however, Fig.~7 clearly show that these data are currently highly consistent 
with each other about $\theta_{13}$, and that their interplay should be described in terms
of synergy rather than tension. Finally, the inclusion of atmospheric data (rightmost panels)
corroborates the previous indications for $\delta$, with a global best fit
around $1.3$--$1.4\pi$ and a slight reduction of the allowed ranges at 1 and 2$\sigma$ (at least for NH). 

Note that, with NO$\nu$A LEM data, the wavy bands in the leftmost panels of Fig.~7 would be
slightly shifted to the right (not shown), leading to slightly stronger bounds on $\delta$
in combination with SBL reactor and atmospheric data, as reported in Fig.~2 of the previous Section; see also the
official NO$\nu$A LID and LEM results in \cite{NOV1}. In this case, one might invoke a slight ``tension''
between LBL accelerator and and SBL reactor data, but only at the level of $\sim 1\sigma$ 
differences on $\theta_{13}$ in the worst case (IH).

Figure~8 shows the $N_\sigma$ bounds in the $(\sin^2\theta_{23},\,\delta)$ plane, which is
gaining increasing attention from several viewpoints, including studies of degeneracies among 
these parameters and $\theta_{13}$ \cite{Mina}, of the interplay between LBL appearance and disappearance channels 
\cite{Park},
and of statistical issues in the interpretation of $N_\sigma$ bounds \cite{Elev}. The bounds in Fig.~8
appear to be rather asymmetric in the two half-ranges of both $\theta_{23}$ and $\delta$, and also quite
different in NH and IH. This is not entirely surprising, since this is the only covariance plot 
(among Figs.~3--8) between two unknowns: 
the $\theta_{23}$ octant (in abscissa) and the CP-violating phase $\delta$ (in ordinate). Therefore,
the contours of Fig.~8 may evolve significantly as more data are 
accumulated, especially by oscillation searches with atmospheric and LBL accelerator experiments.

We conclude this Section by commenting the $\Delta\chi^2_{\mathrm{I-N}}$ values reported in Sec.~3,
which differ only by the inclusion of NO$\nu$A LID data (Table~1) {\em vs\/} LEM data (Table~2). 
In the first case, the $\Delta\chi^2_{\mathrm{I-N}}$ takes the value $-1.2$ for a fit to LBL Acc.+Solar+KL data,
becomes $-0.88$ by including SBL reactor data, and changes (also in sign) to $+0.98$ by including 
atmospheric data.%
\footnote{The latest SK atmospheric data  induce a preference for NH \protect\cite{Wend},
that we also find (but more weakly) in our SK data fit.
}
 Since these $\Delta\chi^2_{\mathrm{I-N}}$ values are both small (at the level of $\pm1\sigma$) and with
unstable sign, we conclude that there is no significant indication about the mass hierarchy,
at least within the global fit including default (LID) NO$\nu$A data.
By replacing NO$\nu$A LID with LEM data (Table~2), the same exercise leads to the following progression of 
$\Delta\chi^2_{\mathrm{I-N}}$ values: $+0.61$ (LBL Acc.+Solar+KL), $+2.2$ (plus SBL Reactor), $+2.8$ (plus Atmospheric).
In this case, a weak hint for NH (at $\sim\!1.6\sigma$, i.e., $\sim\!90\%$ C.L.) seems to emerge from consistent (same-sign) indications 
coming from different data sets, which is the kind of ``coherent'' signals that one would hope to observe,
at least in principle.
Time will tell if these fragile indications about the hierarchy will be strengthened or weakened by future data
with higher statistics.

\section{Implications for absolute neutrino masses}

Let us discuss the implications of the previous oscillation results 
on the three observables sensitive to the (unknown) absolute $\nu$ mass scale: the sum of $\nu$
masses $\Sigma$ (probed by precision  cosmology), the effective $\nu_e$ mass $m_\beta$ (probed by $\beta$ decay),
and the effective Majorana mass $m_{\beta\beta}$ (probed by $0\nu\beta\beta$ decay if neutrinos are Majorana fermions).
Definitions and previous constraints for these observables can be found in \cite{Melc,Prev,Glob}; here we just remark that 
the following discussion is not affected by the current uncertainties on  $\theta_{23}$ or $\delta$.

Figure~9 shows the constraints induced by our global $3\nu$ analysis at $2\sigma$ level, 
for either NH (blue curves) or IH (red curves), 
  in the planes charted by
  any two among the parameters $m_\beta$, $m_{\beta\beta}$ and $\Sigma$.   
 The allowed bands for NH and IH, which practically coincide in the (so-called degenerate) mass region well above $O(10^{-1})$~eV, start to differ significantly at relatively low mass scales of  $O(\sqrt \Delta m^2)$ and below.  At present, $\beta$- and
$0\nu\beta\beta$-decay data probe only the degenerate region of $m_\beta$ and $m_{\beta\beta}$, respectively \cite{PDGR}, 
while cosmological data are deeply probing the sub-eV scale, with upper bounds on $\Sigma$ as low as $\sim 0.1$--0.2~eV,
see e.g.~\cite{Cos1,Cos2,Cos3,Cos4,Cos5,Cos6,Cos7} and references therein. 
Taken at face value, the cosmological bounds would somewhat
disfavor the IH case, which entails $\Sigma$ values necessarily larger than $\sim \!2 \sqrt \Delta m^2 \sim 0.1$~eV (see Fig.~9).
Interestingly,  these indications are consistent with a possible
slight preference for NH from the global $3\nu$ analysis (with NO$\nu$A LEM data), as discussed
at the end of the previous Section. We do not attempt to combine cosmological and oscillation data, but we remark
that the evolution of such hints will be a major issue in neutrino physics for a long time, with challenging implications for $\beta$-decay and $0\nu\beta\beta$-decay searches \cite{ViRe}.

\begin{figure}[t]\centering
  \includegraphics[width=0.53\textwidth]{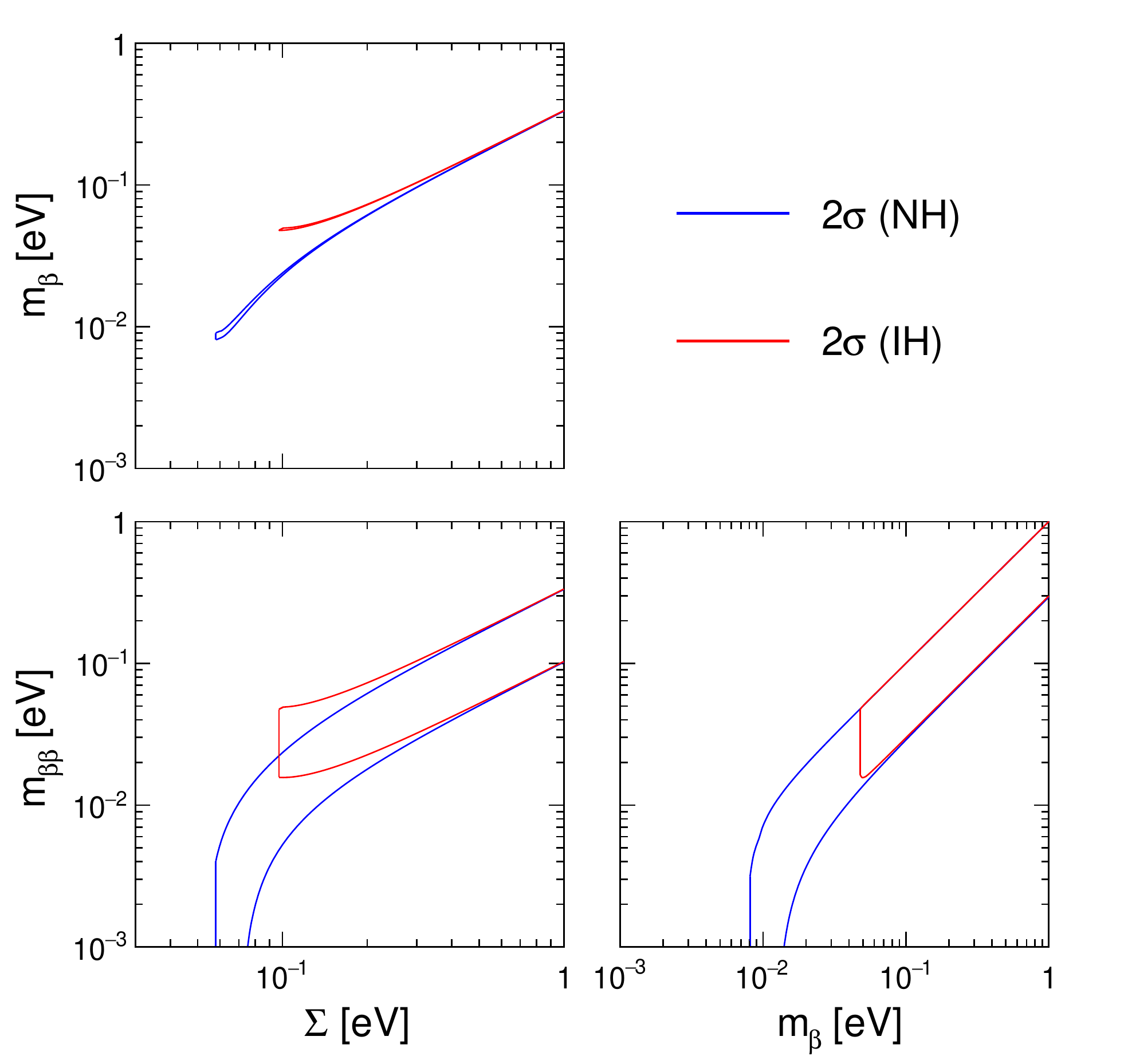}\vspace*{-2mm}
  \caption{\small Constraints induced by the global oscillation data analysis at $2\sigma$ level, for either NH (blue curves) or IH (red curves), 
  in the planes charted by
  any two among the absolute neutrino mass observables $m_\beta$, $m_{\beta\beta}$ and $\Sigma$.
  \label{fig09}}
\end{figure}

\section{Conclusions}

We have presented the results of a state-of-the-art  global analysis of neutrino oscillation data, performed
within the standard $3\nu$
framework.
Relevant new inputs (as of January 2016)
include the latest data from the Super-Kamiokande and IceCube DeepCore atmospheric experiments,
the long-baseline accelerator data from T2K (antineutrino run)  and NO$\nu$A (neutrino run) in both
appearance and disappearance mode, the far/near spectral ratios from the Daya Bay and RENO short-baseline
reactor experiments, and a reanalysis
of KamLAND data in the light of the ``bump'' feature recently observed in reactor antineutrino spectra.

The five known oscillation parameters ($\delta m^2,\,\sin^2\theta_{12},\,|\Delta m^2|,\,\sin^2\theta_{13},\sin^2\theta_{23}$)
have been determined with fractional accuracies as small as (2.4\%,\,5.8\%,\,1.8\%,\,4.7\%,\,9\%), respectively.
With respect to previous fits, the new inputs induce small downward shifts of $\delta m^2$ and $\sin^2\theta_{12}$, and a small increase 
of $|\Delta m^2|$ (see Fig.~1 and Table~1). 

The status of the three unknown oscillation parameters is as follows. The $\theta_{23}$ octant ambiguity remains essentially
unresolved: The central value of $\sin^2\theta_{23}$ is somewhat fragile, and it can flip from the first to the
second octant by changing the data set or the hierarchy. Concerning the CP-violating phase $\delta$, 
we confirm the previous trend favoring $\sin\delta<0$ (with a best fit at $\sin\delta\simeq -0.9$), although
all $\delta$ values are allowed at $3\sigma$. Finally, we find no statistically significant indication in favor of 
one mass hierarchy (either NH or IH).  

Some differences arise by changing the NO$\nu$A appearance data set, from the default (LID)
sample to the alternative (LEM) sample. A few known parameters are slightly altered, as described
in Fig.~2 and Table~2. There is no significant improvement on the octant ambiguity, while the
indications on $\delta$ are strengthened, and some ranges with $\sin\delta>0$ can be
excluded at $3\sigma$ level. Concerning the mass hierarchy, the NH case 
appears to be slightly favored (at $\sim\! 90\%$ C.L.).

We have discussed in detail the parameter covariances and the impact of different data sets
through Figs.~3--8, that allow to appreciate the interplay among the various (known and unknown) parameters, 
as well as the synergy between  oscillation searches in different kinds of experiments.
Finally, we have analyzed the implications of the previous results on the 
non-oscillation observables ($m_\beta,\,m_{\beta\beta},\,\Sigma$) that can probe absolute neutrino masses (Fig.~9).
In this context, tight upper bounds on $\Sigma$ from precision cosmology appear to favor the NH case.  
Further and more accurate data are needed to probe the hierarchy and absolute mass scale of neutrinos, 
their Dirac or Majorana nature and CP-violating properties, 
and the $\theta_{23}$ octant ambiguity, which remain as missing pieces of the $3\nu$ puzzle.

\section*{Acknowledgments}
This work was supported by the Italian Ministero dell'Istruzione, Universit\`a e Ricerca (MIUR) and Istituto Nazionale di Fisica
Nucleare (INFN) through the ``Theoretical Astroparticle Physics'' research projects.

\end{document}